\documentclass[12pt]{JHEP3}
\pdfoutput=1
\usepackage{amsmath, amsfonts,euscript,bbm}
\usepackage{ifthen}
\usepackage{graphicx}
\usepackage{psfrag}

\def\gsim{\mathrel {\vcenter {\baselineskip 0pt \kern 0pt \hbox{$>$} \kern 0pt \hbox{$\sim$} }}}
\let\d=\delta
\let\e=\epsilon

\let\P=\Pi

\def\pa{\partial}
\def\P{\mathcal{P}}
\def\dc{\delta \chi}

\newcommand{\PRE}[1]{}
\newcommand{\Expect}[1]{\left\langle #1 \right\rangle}

\newcommand{\be}{\begin{equation}}
\newcommand{\ee}{\end{equation}}
\newcommand{\bea}{\begin{eqnarray}}
\newcommand{\eea}{\end{eqnarray}}
\newcommand{\beas}{\begin{eqnarray*}}
\newcommand{\eeas}{\end{eqnarray*}}

\newcommand{\nbox}{{\,\lower0.9pt\vbox{\hrule \hbox{\vrule height 0.2 cm
\hskip 0.2 cm \vrule height 0.2 cm}\hrule}\,}}

\def\lsim{\mathrel {\vcenter {\baselineskip 0pt \kern 0pt \hbox{$<$} \kern 0pt \hbox{$\sim$} }}}

\newcommand{\V}[1]{\vec{#1}}

\title{Scale Dependent Local Non-Gaussianity from Loops}

\author{
Jason Kumar${}^{1}$, Louis Leblond${}^{2,3}$, Arvind Rajaraman${}^{4}$\\
${}^{1}$Department of Physics and Astronomy, University of
Hawaii, \\ \qquad \qquad Honolulu, HI  96822, USA\\
${}^{2}$George P. \& Cynthia W. Mitchell Institute for Fundamental
Physics,  \\ \qquad \qquad Texas A\&M University, College Station, TX
77843, USA\\
${}^{3}$Perimeter Institute, 31 Caroline St, Waterloo, On, N2L 2Y5, Canada\\
${}^{4}$Department of Physics and Astronomy, University of
California, \\ \qquad \qquad  Irvine, CA 92697, USA
}
\date{}

\abstract{We analyze multi-field inflationary systems which yield
strongly scale dependent non-Gaussianity with a shape that is very close to the local shape.
As in usual multi-field models, the non-Gaussianity arises from the non-linear transfer of
scalar field fluctuations to curvature perturbations.  Here we consider models in which
higher order terms (loops) dominate over the lowest order source of non-linearity.
The magnitude of non-Gaussianity depends on an infrared cutoff which
is determined by our observational probes measuring
non-Gaussianity.
In our models, the running
is positive and large ($n_{NG} \sim 0.2$) on CMB scales.
The magnitude of the bispectrum is maximally of order $\mathcal{O}(100)$,
and grows on small scales. This can lead to interesting signals for
large scale structure.}

\preprint{UCI-TR-2009-15, UH511-1141-09, MIFP-09-37, arXiv:0909.2040}

\keywords{Effective Field Theory, Cosmology, Inflation}

\begin{document}

\section{Introduction}

With the advent of precise cosmological data,
it is now possible to constrain models of inflation by the measured magnitude
and scale-dependence of correlated temperature perturbations in the cosmic microwave background (CMB)
and from tracking density perturbations in dark matter from measuring the Large Scale Structure (LSS) of
our universe. In these observations, it is found that the primordial perturbations coming from
inflation are
Gaussian to a remarkable accuracy, in agreement with the  predictions of most
single field
models of inflation.

Non-Gaussianity (NG) can  be quantified by the magnitude of the bispectrum denoted $f_{NL}$
(this is usually quoted at the equilateral point in momentum space where all three momenta are equal).
For most slow-roll models, $f_{NL}$ is smaller than 1~\cite{Maldacena:2002vr, Acquaviva:2002ud}.
By comparison, the most recent constraints from WMAP5 \cite{Komatsu:2008hk} data are
$-4<f_{NL}<80$ for the local shape and $-125<f_{NL}^{equi}<435$ for the equilateral shape
\cite{Senatore:2009gt}.
The Planck satellite is expected to improve the bounds to  $\Delta f_{NL}  < 7$ \cite{Cooray:2008xz}.
There are also a large number of running and upcoming experiments probing LSS scales (such as LSST, DES, SDSS, etc.)
and they may allow us to eventually probe non-Gaussianity on smaller scales.

In this note, we shall consider multi-field models with a large bispectrum
(three-point correlation function) that is strongly scale dependent\footnote{There has been much recent work in calculating the bispectrum and trispectrum in multi-field inflation, for some recent references see \cite{Langlois:2008vk, Gao:2009at, Arroja:2008yy, Huang:2009xa, Huang:2009vk, Byrnes:2009qy, Battefeld:2009ym}.}. The running is positive (or blue which 
means that the NG grows as $k$ increases)
and can be achieved while keeping the power spectrum nearly scale invariant.
It arises from loops (or higher order terms in the local ansatz) and the shape of the bispectrum
is very well approximated by the local shape multiplied by a logarithm. We
provide a consistent setup where the 1-loop effect dominates the bispectrum while giving a subdominant contribution
to the power spectrum, and where higher loop contributions can be neglected.
Since the running is positive, we can engineer a
set-up where the curvature perturbation on CMB scales are extremely
Gaussian while having a detectable NG on LSS scales. 

Running NG has already been considered in the context of DBI inflation \cite{Alishahiha:2004eh, Silverstein:2003hf}. 
This model can have a strong 
NG signal due to a small and varying sound speed for the inflaton 
fluctuations~\cite{Chen:2005fe}. The amplitude of the 3-pt can strongly run with 
scale if the sound speed varies but the running of the sound speed is exactly 
cancelled by the quickly varying Hubble constant along the trajectory. This is the 
key point of this type of model where the potential is steep but the inflaton moves 
slowly because of a speed limit. This causes the power spectrum to be scale invariant 
while the bispectrum can run wildly \cite{ArmendarizPicon:2003ht, Khoury:2008wj}.

The prospect of detecting large NG with large
scale structure data has spurred much activity recently. LoVerde
et al \cite{LoVerde:2007ri} have examined the possibility of using cluster
counts and the galaxy bispectrum to constrain running $f_{NL}$.
It was also realized in \cite{Dalal:2007cu, Matarrese:2008nc},
that NG of the local shape can induce a scale dependence of the galaxy/halo bias
(see also \cite{Slosar:2008hx, Afshordi:2008ru, McDonald:2008sc, Seljak:2008xr,Taruya:2008pg, Grossi:2009an}).
 This effect can be easily found in the data and it results in a competitive bound on
NG with local shape $ -29 < f_{NL} < 70$  \cite{Slosar:2008hx}.
At the time of this writing, there exists no significant  experimental
bound on the running of NG with scale. Recently, Sefusatti et
al \cite{Sefusatti:2009xu} argued that Planck could bound $n_{NG}$, the running
of non-Gaussianity, with a precision $\Delta n_{NG} \sim 0.1 $ $(0.3)$ for a
local (equilateral) shape of non-Gaussianity.

In our models, we find NG with a (nearly) local shape with a scale dependence
such that the NG signal grows on small scales.
The magnitude of the bispectrum grows with $k$ with a model independent running of $n_{NG} \sim 0.2$ at
CMB scale and $0.1$ on LSS scale. The strongest constraint on the magnitude of NG arises
from $n_s$. We find that $f_{NL} \sim 100$ can be achieved in principle.
We also calculate the trispectrum $\tau_{NL}$, which also runs. 
Before getting into the details, we summarize the basic idea and results.

\section{Scale Dependence from Loops}
Local shape NG can be obtained in multi-field models of inflation, where each field is
Gaussian but a non-linear relation between the inflaton perturbations and curvature
perturbations induces NG.  The original definition of the local ansatz for the curvature perturbation
was done in real space \cite{Komatsu:2001rj}
\begin{equation}\label{localansatz}
\zeta(\vec{x},t) = \zeta_{Gauss} + \frac{3}{5}f_{NL} (\zeta_{Gauss}^2 - \Expect{\zeta_{Gauss}^2})\; ,
\end{equation}
where $\zeta_{Gauss}$ is the Gaussian piece of the curvature perturbation. $f_{NL}$ in this
formula is by definition scale invariant. In momentum space,
the above ansatz leads to the following bispectrum
\bea
\label{fnldef2}
\Expect{\zeta_{\vec{k}_1} \zeta_{\vec{k}_2}\zeta_{\vec{k}_3}} & = & \frac{3}{5} f_{NL}
\Expect{\zeta_{\vec{k}_1} \zeta_{\vec{k}_2}(\zeta \star \zeta)_{\vec{k}_3}}\nonumber\\
&=& (2\pi)^7\delta^3(\sum\vec{k}_i) \frac{3}{10} f_{NL} (\mathcal{P}^\zeta)^2 \frac{\sum k_i^3}
{\prod k_i^3}\; ,
\eea
where $(\zeta \star \zeta)_{\vec{k}_3}$ denotes a convolution, $\mathcal{P}^\zeta$ is the power
spectrum (which is assumed to be scale invariant, for simplicity)
and   $\frac{\sum k_i^3}{\prod k_i^3}$ defines
the local shape.
Many multi-field models (such as curvatons \cite{Linde:1996gt, Lyth:2001nq})
have local scale invariant NG of this type. The NG can also be scale dependent
even if the shape is nearly local; for example, this is expected to happen when the NG is
generated throughout the whole trajectory as opposed to simply at some fixed later time, such as in
curvaton models. A particular model with this feature was considered by Byrnes et al \cite{Byrnes:2008zy, Byrnes:2008wi}, where
the scale-dependence arises from the dependence of $f_{NL}$ on the (time-dependent)
slow-roll and Hubble parameters. In their case, the NG decreases on small scales.

We instead look for scale dependence coming from loops and higher order terms. Indeed,
it was realized early on \cite{Lyth:2005fi} that an additional contribution to the bispectrum in
the ansatz Eq.~(\ref{localansatz}) comes from
\be
\Expect{\zeta_{\vec{k}_1} \zeta_{\vec{k}_2}\zeta_{\vec{k}_3}} =  \left(\frac{3}{5}
f_{NL}\right)^3\Expect{(\zeta \star \zeta)_{\vec{k}_1}(\zeta \star \zeta)_{\vec{k}_2}
(\zeta \star \zeta)_{\vec{k}_3}} \; .
\ee
This higher order contribution to the bispectrum has a structure similar from a loop contribution as it 
involves an integral over internal momenta. The integral converges in the UV but contains
IR divergences if the power spectrum is nearly scale invariant. One can `regulate' this divergence by
introducing an IR cutoff in momenta $1/L$ \footnote{These loops have been
called c-loops~\cite{Lyth:2006qz}. They must not be confused with q-loops, or loops coming from
the expansion of the quantum evolution operator
prior to horizon crossing \cite{Weinberg:2005vy}. There has been much discussion recently on the
physical significance of the IR divergences in loop calculation in inflation.
For c-loops, this IR cutoff is physical and depends on the
observational probe and on how we measure the zero mode of curvature perturbations. We will justify
this point of view in more detail in Sec.~(\ref{IR}).}.
Doing so, the shape of this term is close to local up to a log  \cite{Lyth:2005fi, Boubekeur:2005fj}
\be
\Expect{\zeta_{\vec{k}_1} \zeta_{\vec{k}_2}\zeta_{\vec{k}_3}} \propto \ln(\rm{Min}[k_i] L)
\frac{\sum k_i^3}{\prod k_i^3}\; .
\ee
If this term dominates the bispectrum, we will have a scale dependence with a running of order $n_{NG}
\sim \frac{1}{\ln kL}$.
As we will show later, the cutoff $L$ is well approximated by the size of the universe today such that
$\ln kL \sim 5$ around CMB scale and $n_{NG} \sim 0.2$. The NG grows with scale becoming more important
for smaller wavelength. Needless to say this is the interesting case as it gives rise to a stronger signal
for LSS.

Recently, Cogollo et al  \cite{Cogollo:2008bi} and Rodriguez et al \cite{Rodriguez:2008hy}
have argued that loops can dominate in a particular 2-brid model. While their idea
is very similar to what we propose, their particular model suffers from a problem
pointed out in \cite{Byrnes:2008zy}. One of the fields that is assumed to follow a
smooth classical trajectory is actually dominated by its quantum fluctuations,
undermining part of their analysis.

As we will show, the field that gives rise to NG in our model is also dominated
by its quantum fluctuations.  But this field plays no role in the inflationary
trajectory and there is no inconsistency.
We consider multi-field models of hybrid inflation
where the inflationary trajectory is dictated by a
single field but
the surface of reheating (determined by when an extra waterfall/tachyon field starts condensing)
fluctuates due to two fields \cite{Dutta:2008if, Dutta:2007cr}
(as originally envisioned by \cite{Alabidi:2006wa, Alabidi:2006hg}  -- see
also \cite{Bernardeau:2007xi, Sasaki:2008uc, Naruko:2008sq} for similar models).

 In section 3, we describe the detailed set-up for the model, and describe the
infra-red momentum cutoff.  In section 4 we compute the power spectrum, and in
section 5 we compute the bispectrum and trispectrum.  We conclude in section
6 with a discussion of these results.

\section{Multi-Field Model}

\begin{figure}[ht]
\centering
\includegraphics[width=3.5in]{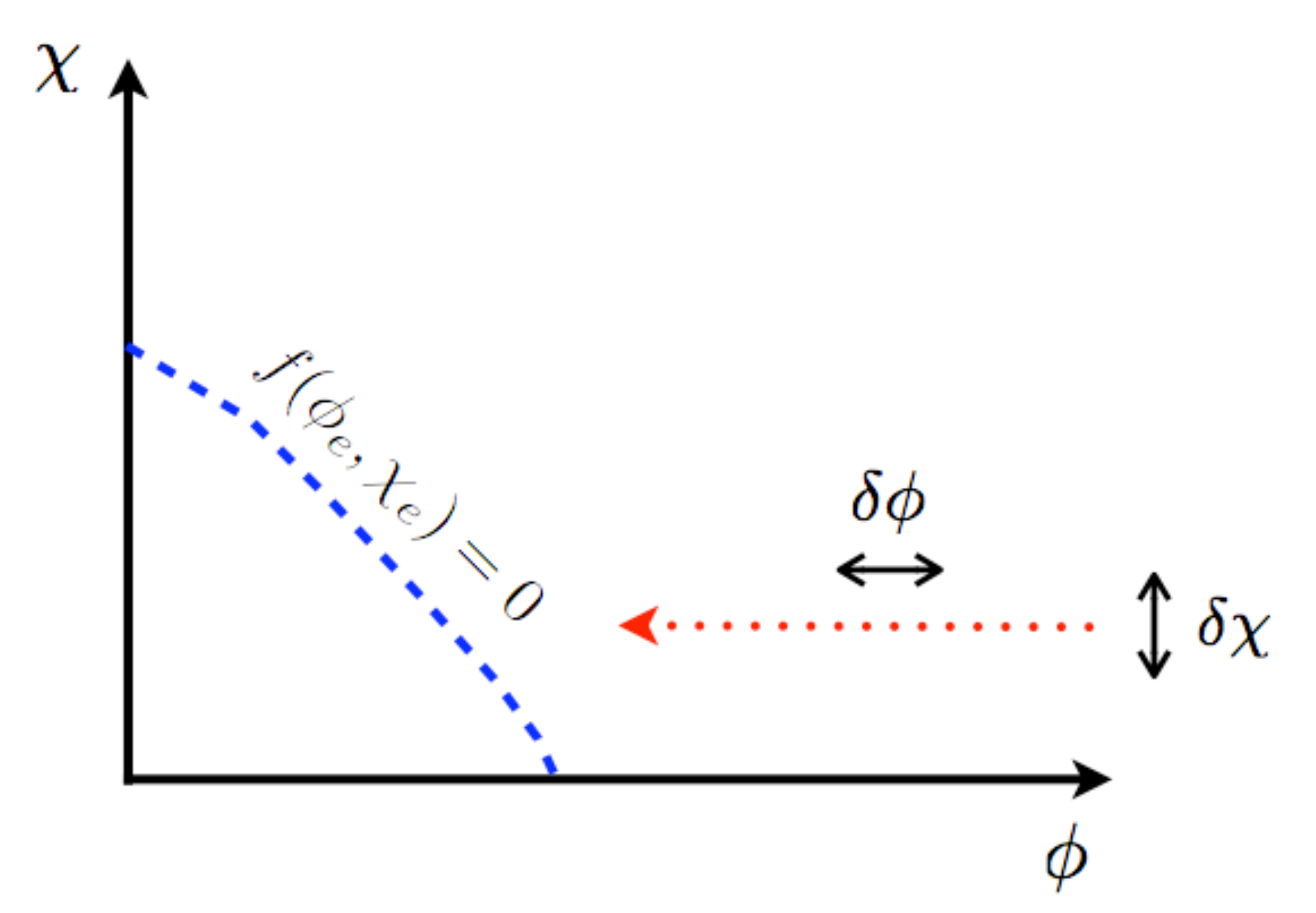}
\caption{This figure depicts the trajectory in field space. The blue (dashed) line denote the surface of
reheating defined by $f(\phi_e,\chi_e)=0$ and it is assumed to be thin. The classical trajectory
is in the $\phi$ direction (red/dotted line) but both $\delta\phi$ and $\delta\chi$ will induce curvature
perturbations.}\label{surfacereheat}
\end{figure}

A simple way to move beyond single field slow-roll and generate NG is to have multiple fields.
This type of model can quickly become very complicated and in order to simply illustrate
the main physical effect of interest (namely large scale dependent NG from loops), we will
consider a very simplified set-up. More general models and in-depth analysis of the model
we present is left for future work.
Consider a model of hybrid inflation with two real light scalar fields ($\phi$ and $\chi$) and a
waterfall field $T$ which ends inflation when it becomes tachyonic and condenses. In this paper, we
will consider a rather general action, a more detailed and worked example is given
in Appendix \ref{details}. The action
is (we follow the notation of~\cite{Dutta:2007cr}):
\bea
S & = &\frac12 \int\sqrt{g}[M_p^2 R - (\partial\phi)^2- (\partial T)^2 - (\partial\chi)^2 -2 V]\; ,
\nonumber\\
V &=& V_{\rm{inf}}(\phi) + V_{\rm{hid}}(\chi) + V_{\rm{mess}}(\phi,\chi,T)]\; .
\eea
The only coupling between $\phi$ and $\chi$ are through the tachyon which acts as a mediator or
messenger. The form of $V_{\rm{mess}}$ is taken to be
\be\label{Vmess}
 V_{\rm{mess}} \propto T^2 f(\phi, \chi) + \mathcal{O}(T^n) \;\;\; ; n>2 \; .
\ee
The function $f$ interpolates from large and positive values (in Hubble units) during inflation
to negative values after the system crosses a
critical line in field space. Therefore during inflation, $T$ has
a large positive mass, its vev is driven to zero and its potential vanishes. Because of its large
mass, this field will not fluctuate and it can be integrated out of the theory.  In this model,
inflation ends suddenly when the mass of the tachyon vanishes, which occurs on a line in
field space parameterized by
\be\label{SurfaceReheat}
f(\phi_e,\chi_e) = 0 \; ,
\ee
where the index ``$e$" denotes the value of the fields at the end of inflation.
During the inflationary phase, $\phi$ and $\chi$ have no direct coupling. To simplify
further, we assume that $V_{hid}(\chi) \ll V_{inf}(\phi)$ and we refer to $\phi$ as
the inflaton from now on. The Hubble scale is then approximately given by
\be
 H^2  \approx \frac{V_{\rm{inf}}}{3M_p^2}
\ee
and $\chi$ is a ``hidden" field during inflation which fluctuates but without much impact on
the total energy density of the Universe.
Nevertheless, its quantum fluctuations are still important as they will be felt  as
ripples on the surface of reheating. Indeed, at different point in space, the (slightly)
different value of $\chi_e$ will mean different critical value $\phi_e$ for the inflaton
resulting in more or less inflation in these different regions. This correlates directly
in curvature perturbations (See Fig.~(\ref{surfacereheat}))
Since the quantum perturbations of $\chi$ mainly affect
the surface of reheating, this system is well amenable to analysis through the $\delta N$
(or separable universe) formalism~\cite{Sasaki:1995aw}. The idea is that the
curvature perturbation on large scales is simply given by the perturbation in the number of
efolds for each trajectories
\be\label{deltaN}
\zeta(\vec{x},t) = \delta N(\vec{x},t)\; ,
\ee
where the curvature perturbation $\zeta$ is given by fluctuations of
the scale factor $a(\vec{x},t) = a(t) e^{\zeta(\vec{x},t)}$ and the
difference in number of efolds is from a initial flat hypersurface to a uniform energy
density final hypersurface. This formula does not take into account possible interactions
between the various fields inside the horizon (on small scale) and it is only valid after
horizon crossing where the evolution of the curvature perturbation is
classical\footnote{The $\delta N$ formalism will not account correctly for multi-field
effects for modes inside the horizon. In our case, because the fields are uncoupled during inflation,
we can solve for $\delta\phi$ and $\delta\chi$ are horizon exit independently and follow
the subsequent evolution of $\zeta$ with the $\delta N$ formalism.}.

The surface where inflation ends Eq.~(\ref{SurfaceReheat}) is not a uniform energy
density hypersurface and a correction term must be included as discussed
in \cite{Vernizzi:2006ve, Sasaki:2008uc}. The correction term is very small in
the hybrid scenario where the potential is very flat and it will be dropped in what
follows. The number of efolds is given by $dN = -Hdt$. For the case where the
classical trajectory is determined by a single field $\phi$, one has
\be
N = - \int_{\phi_*}^{\phi_e(\chi)} \frac{H}{\dot\phi}d\phi^\prime \; ,
\ee
where the critical value of $\phi$ depends on the value of the field $\chi$ at the end of
inflation (we dropped the subscript $e$ and $\chi = \chi_e$ unless otherwise
specified\footnote{The field $\chi$ is evolving stochastically and the value of the
field at the end of inflation is the sum of all fluctuations created for each mode as they
exit the horizon.}) and $*$ refers
to horizon crossing for a given mode.
By varying $\phi_* \rightarrow \phi_* + \delta\phi$ and then  $\phi_e(\overline{\chi}+\delta\chi) = \phi_e
+ \gamma\delta\chi + \gamma_{,\chi}\delta\chi^2/2 + \cdots$ with
\be
\gamma(\overline\chi) = \frac{\partial\phi_e}{\partial\overline\chi}
\ee
where we denote the zero mode of $\chi$ by $\overline\chi$, that is $\chi(\vec{x},t) = \overline\chi(t) 
+ \delta\chi(\vec{x},t)$ (for notational simplicity, the bar is omitted in any derivative subscript).  
We get at second order (using $\frac{H}{\dot\phi} = - N^\prime$)
\be\label{eqN}
\delta N = N^\prime\delta\phi\big|_* - N^\prime \gamma \delta\chi\big|_e
+ \frac12N^{\prime\prime}\delta\phi^2\big|_* - \frac12N^\prime\gamma_{,\chi}\delta\chi^2\big|_e
- \frac12N^{\prime\prime}\gamma^2\delta\chi^2\big|_e \; ,
\ee
where $'$ denotes derivatives with respect to $\phi$. This can be reproduced using the
formula of Vernizzi and Wands \cite{Vernizzi:2006ve}, for the case $\epsilon^\chi \ll \epsilon^\phi$
albeit they implicitly assume that all fields obey their equation
of motion which is not true here for the field $\chi$. It is simple to show that
$N' = \partial N / \partial \phi =1/\sqrt{2\epsilon^\phi} M_p$ where the slow-roll parameters are
\begin{align}
\e^\phi &= \frac12 M_p^2 \left(\frac{V_{\rm{inf},\phi}}{V}\right)^2   \; ,& \e^\chi &=
\frac12 M_p^2 \left(\frac{V_{\rm{hid},\chi}}{V}\right)^2 \; . 
\end{align}
The terms with $N''$ involve derivatives of slow-roll parameters and will therefore be suppressed.
To simplify the formula and the analysis we will consider the case where the slow-roll parameter
at horizon crossing and at the end are equal, $\e_e^\phi = \e_*^\phi$. This is not true in many
models and we will discuss at the end how that would affect our results. We thus drop
all subscript referring to the time of evaluation. The mean of Eq.~(\ref{eqN}) is non-zero and as
it is we will generate a one-pt function. To ensure that the mean is zero we can subtract a
constant piece (keeping only the leading terms)
\bea\label{curveRealSpace}
\zeta & = &N^\prime\delta\phi - N^\prime \gamma \delta\chi - \frac12N^\prime\gamma_{,\chi}
\delta\chi^2 + \frac12N^\prime\gamma_{,\chi} \Expect{\delta\chi^2}\; ,
\eea
which is of the form Eq.~(\ref{localansatz}).
\be
\zeta =  \zeta_1 + \zeta_2 - \Expect{\zeta_2} \; .
\ee
Note that this series terminates if
\begin{enumerate}
\item the function $\gamma$ is such that $\gamma_{,\chi\chi}$ and higher derivatives are small.
\item $N''$ and higher derivative contributions are small.
\end{enumerate}
In this type of model, the function $\gamma$ could be anything and in the case
where $\gamma_{,\chi}\delta\chi > \gamma$ the quadratic piece in $\delta\chi$ will dominate over
the linear piece (in $\delta\chi$) which ensures that the loop contribution to the bispectrum will
dominate
\be
\Expect{\zeta^3} \propto \gamma_{,\chi}^3 \Expect{(\delta\chi^2)^3}\; ,
\ee
as we advocated earlier. In order for the power spectrum to be nearly scale invariant we
will still need the $\delta\phi$ piece to be the dominant contribution to the power
spectrum.  There is no
contradiction since the linear perturbation in $\phi$ does not contribute to the bispectrum
(or gives a very small
slow-roll suppressed contribution). Furthermore, in the case where the
higher derivatives of $\gamma$ are suppressed, the higher loop contribution can be neglected,
ensuring a consistent truncation.

Another important point is that for the loop to dominate, the zero mode of $\chi$ at the end of 
inflation ($\overline\chi_e$ which is the mean averaged over the size of the universe at the end 
of inflation) must be smaller then the 1-$\sigma$ deviation value of the perturbation around the mean. 
Taking the quantum perturbation to be of order $\delta\chi \sim H$, we must 
have $\overline\chi_e < \delta\chi$. This is better seen in a specific model such as the one 
presented in Appendix \ref{details}. There, we use a model where $\phi_e = f(\chi^2)$ such 
that $\gamma \propto \chi$ and $\gamma_{,\chi} \sim \rm{cst}$ and the series truncate. In such 
models it is clear that the quadratic term dominate over the linear piece when
\be
\frac{\gamma_{,\chi}\delta\chi}{\gamma} \sim \frac{\delta\chi}{\chi} > 1\; .
\ee
It is then clear that the field $\chi$ has to behave stochastically and is in no way 
following a classical equation of motion.  The fact that $\chi$ has essentially no effect on 
the inflationary dynamics prior to the reheating tells us that the stochastic behavior is 
unimportant during inflation. The value of $\overline\chi_e$, being stochastic, could have any 
value and it is therefore a free parameter.  
Before going into more details of the calculation, we 
need to discuss the choice of IR cutoff in the loop calculation.

\subsection{The IR cutoff}
\label{IR}

There has been much discussion in the literature
about the choice of cutoff that should be used in loop calculations.
For the calculation of quantum loops in the in-in formalism (prior to horizon exit), the
correlations function of scalars appear to be sensitive to this choice of
cutoff, and there is no clear understanding of how this cutoff should be set.
But for the c-loops which we consider in this paper, the situation is considerably
simpler and there is a natural choice of cutoff~\cite{Lyth:2007jh}.
We will define the observed zero modes of the fields $\phi,\chi$ as
\bea
\phi_0={1\over L^3}\int_{-L/2}^{L/2}d^3x \phi\; ,\qquad \qquad \chi_0={1\over L^3}
\int_{-L/2}^{L/2}d^3x \, \chi \; ,
\eea
where $L$ is the largest scale over which we have measured the fields.
The perturbations of the fields are then defined as $\delta\phi=\phi-\phi_0,
\delta\chi=\chi-\chi_0$.

When computing correlators of $\d N$, we are actually interested in the
correlations functions of the perturbations e.g. $\langle \d\phi_{\V{k}_1} \d\phi_{\V{k}_2} \rangle$.
From the definition of the perturbations, we see that the effect of subtracting the zero mode is to
remove all Fourier modes with momentum $k>L^{-1}$. Hence $\d\phi_{\V{k}}=\phi_{\V{k}}$ for $k>L^{-1}$,
and zero otherwise.
Similarly, we find
\bea
\langle \d\phi_{\V{k}_1} \d\phi_{\V{k}_2} \rangle=\left\{ \begin{array}{c}(2\pi)^3 \delta^3(\vec{k}_1 +
\vec{k}_2)
2\pi^2 {\mathcal{P}_*\over k_1^3}\qquad k>L^{-1}
\\
0\qquad k <L^{-1}\end{array}\right. \; .
\eea
The effect is to include a cutoff $L^{-1}$ on any momentum integral.
Due to the cutoff, the correlation functions will have an explicit dependence
on $L$. This can be traced back directly to the fact that we are
calculating correlation functions of perturbations like $\d\phi=\phi-\phi_0$, which have
a direct dependence on $L$ through $\phi_0$.
In this formalism, it is clear that all the dependence on $L$
comes from the variation in the zero mode as a function of $L$ as was discussed
in more details in~\cite{Lyth:2007jh} (see also \cite{Enqvist:2008kt}).

To summarize, there is a natural cutoff $L$ determined by the biggest scale on which we are able to measure
the background zero mode of curvature.  This is maximally  the size of the universe today $L\sim 1/H_0$.
This coincides with the lowest $k$ perturbations that are possible to observe now. Since there are
about 5 efolds between when the lowest observable wavenumber leaves the horizon and when CMB scales
leave the horizon, we have $k_{CMB} L\sim e^5$.
LSS are about
two orders of magnitude greater than CMB scales, giving $k_{LSS} L \sim e^{10}$.

\section{The Power Spectrum}

We will first consider the two-point function
$\langle \zeta_{k_1} \zeta_{k_2} \rangle$.  For the scalar fields, we have
\bea
\Expect{\delta\chi^2_{\V{k}}}  &= & \Expect{\delta\phi^2_{\V{k}}} =  (2\pi)^3 \delta^3(\sum_i \vec{k}_i)
P(k)\; ,\nonumber\\
P(k) & = & \frac{2\pi^2\mathcal{P}}{{k^3}}\; ,
\eea
and we consider a model where these expectation values are approximately constant
and where $\mathcal{P}$ is scale invariant
(independent of $k$). We will also assume that any intrinsic 3-pt functions are negligible,
$\Expect{(\d\phi_k)^n}\approx 0$ and $\Expect{(\d\chi_k)^n}\approx 0$ for $n$ odd. In Fourier space
the curvature perturbation is given by (from Eq.~(\ref{curveRealSpace}))
\be
\zeta_{\V{k}}  = N^\prime\delta\phi_{\V{k}} - N^\prime \gamma \delta\chi_{\V{k}} - \frac12N^\prime
\gamma_{,\chi}
\int \frac{d^3\V{k}'}{(2\pi)^3}\delta\chi_{\V{k}-\V{k}'}\delta\chi_{\V{k}'} + \frac12N^\prime
\gamma_{,\chi} \Expect{\delta\chi_{\V{k}}^2}\; ,
\ee
The ``tree-level"
contribution to the power spectrum arises from
linear terms in the expansion of $\delta N$, and it is easily
seen to give
\bea
\Expect{\zeta^2_{\V{k}}}_{tree} &= &N'^2 (\Expect{\delta\phi^2_{\V{k}}}
+ \gamma^2 \Expect{\delta \chi^2_k} )\; ,\nonumber\\
&= &N'^2 (1 + \gamma^2)(2\pi)^3 \delta^3(\sum_i \vec{k}_i) P(k)\; .
\eea
However, there is also a ``one-loop" contribution which
arises from the non-linear terms in the $\delta N$ expansion which leads to
\bea
\Expect{\zeta_{\V{k}_1}\zeta_{\V{k}_2}}_{loop} & = & N'^2\frac{\gamma_{,\chi}^2}{4}
\int \frac{d^3\V{k}'}{(2\pi)^3}\frac{d^3\V{k}''}{(2\pi)^3}
\Expect{\delta \chi_{\V{k}_1-\V{k}'}\delta \chi_{\V{k}'}
\delta \chi_{\V{k}_2-\V{k}''}\delta \chi_{\V{k}''}}\; ,
\nonumber\\
& = & N'^2\frac{\gamma_{,\chi}^2}{4}   (2\pi)^3 \delta^3(\sum_i \vec{k}_i)
\int \frac{d^3k'}{(2\pi)^3}
 \frac{(2) (2\pi^2 \P)^2}{|\vec{k} - \vec{k'}|^3k'^3}\; ,
\eea
where the factor of $(2)$ is from the combinatorics. For a scale invariant power spectra $\mathcal{P}$,
the integral is approximately
\be
\int_{1/L} ^{k}  \frac{d^3\V{k}'}{(2\pi)^3} \frac{1}{|\vec{k}-\vec{k'}|^3k'^3}\; ,
\ee
where we use $k$ as the upper limit because for $k' > k$
the denominator goes as $k'^n$ with $n>3$,
and the integrand drops rapidly. The integrand has two simple poles which give logarithmic divergences.
We regulate these by putting an IR cutoff on the integral.  Hence for this example, we get
\be
\int_{1/L} ^{k}  \frac{d^3\V{k}'}{(2\pi)^3} \frac{1}{|\vec{k}-\vec{k'}|^3k'^3} \sim 2 \frac{\ln(kL)}
{2\pi^2}\; .
\ee
This contribution will depend on the IR limit of the
momentum integration.  This limit is given by the size
of the observable universe today, $L\sim H_0^{-1}$ as we discussed in Sec.~(\ref{IR}).  Modes of longer
wavelength
are already summed in the background value of the
field.  We thus find
\bea
\Expect{\zeta^2_{\V{k}}}_{loop} & = &
N'^2\gamma_{,\chi}^2  (2\pi)^3 \delta^3(\sum_i \vec{k}_i)\frac{2\pi^2\P^2 \ln(kL)}{k^3}\; .
\eea
Combining these terms yields
\bea
\label{powerspectrum}
\Expect{\zeta^2_k} & = & (2\pi)^3 \delta^3(\sum_i \vec{k}_i)\frac{2\pi^2\mathcal{P}^\zeta}{{k^3}}\; , \\
&= & N'^2 (2\pi)^3 \delta^3(\sum_i \vec{k}_i)P
\left[ 1 + \gamma^2 +\gamma_{,\chi}^2 \P \ln(kL)\right]\; .
\eea
We have defined the power spectrum for curvature with the superscript $\zeta$. The
spectral index $n_s - 1 = \frac{d\ln \mathcal{P}^\zeta}{d\ln k}$ is
\bea
n_s -1 &=& {\gamma_{,\chi}^2 \P \over
{1+\gamma^2 + \gamma_{,\chi}^2 \P \ln kL }}\; .
\eea
Note that the log contribution is positive (blue) and if this is the only contribution, we cannot
match to the currently observed value of $n_s \sim 0.96$ \cite{Komatsu:2008hk}. For now,
we simply impose that the
log contribution contribute no more than a percent correction to $n_s$
\be
\gamma_{,\chi}^2 \P \lsim 10^{-2}\; ,
\ee
which in turn implies that the non-linear contribution to the
2-point function must be subleading if $\log (kL) \sim 1$.

\section{Higher Point Functions}

\subsection{Bispectrum}\label{bispectrum}
\begin{figure}[ht]
\centering
\includegraphics[width=5in]{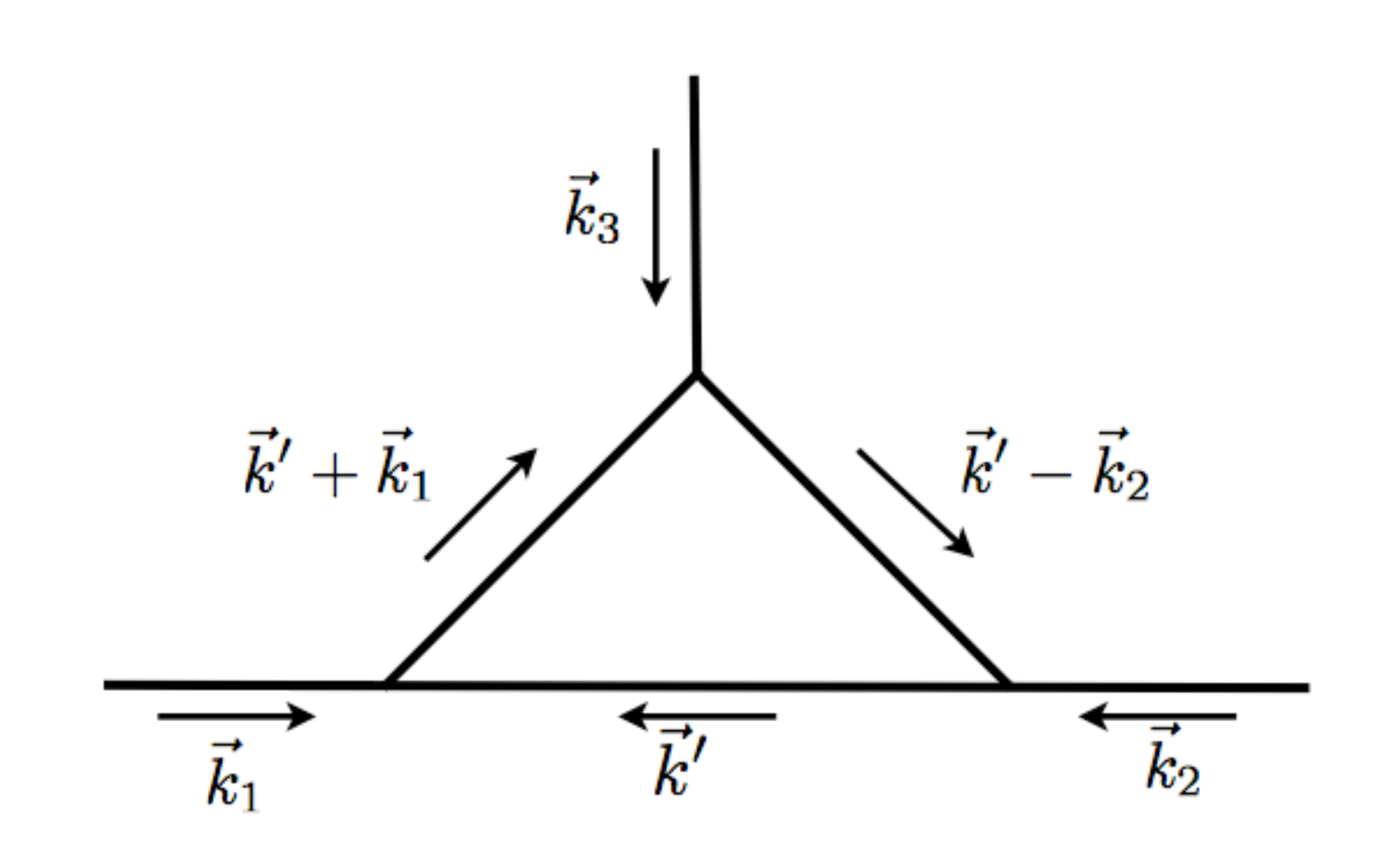}
\caption{The 1-loop diagram. In our case, each vertex is accompanied by a factor of $N'^3\gamma_{,\chi}^3$
while each internal propagator is given by $\frac{2\pi^2\mathcal{P}}{p^3}$. More detailed Feynman rules
for use with the $\delta N$ expansion (which we are not carefully describing here) can be found in
\cite{Byrnes:2007tm}.}\label{tria}
\end{figure}

We  now  compute the 3-point function
$\langle \zeta_{\V{k}_1} \zeta_{\V{k}_2} \zeta_{\V{k}_3} \rangle$.
Again, we find that this correlation function can
easily be computed by expanding $\delta N $ in terms of
$\delta \phi$ and $\delta \chi$.  Since $\delta \phi$ and
$\delta \chi$ are Gaussian fields, the only non-trivial
contributions will come from non-linearities in the
$\delta N$ expansion.
As in the case of the 2-point function, there is
a natural separation into ``tree-level" and ``loop"
contributions~\cite{Zaballa:2006pv}.  The contribution which is of lowest
order in $\gamma_{,\chi}$ is
\bea\label{bitree}
\langle \zeta_{\V{k}_1} \zeta_{\V{k}_2} \zeta_{\V{k}_3} \rangle_{tree} &=&
-\gamma^3 N'^3
{1\over 2}{\gamma_{,\chi} \over \gamma} (3) \int { d^3 \V{k}' \over (2\pi)^3}
\langle \delta \chi_{\V{k}_1} \delta \chi_{\V{k}_2}
\delta \chi_{\V{k}_3-\V{k}'} \delta \chi_{k'}\rangle\; ,
\nonumber\\
&=&-\gamma^3 N'^3
(2\pi)^3 \delta^3 (\sum_i \vec{k}_i)
{\gamma_{,\chi} \over \gamma}
{(2\pi^2\P)^2} \frac{\sum_i k_i^3}{\prod_ik_i^3}\; .
\eea
The next term in the $\gamma_{,\chi}$ expansion is
\bea
\langle \zeta_{\V{k}_1} \zeta_{\V{k}_2} \zeta_{\V{k}_3} \rangle_{loop} &=&
-\gamma^3 N'^3
{1\over 8}{\gamma_{,\chi}^3 \over \gamma^3}
\int \frac{d^3 \V{k}'d^3 \V{k}''d^3 \V{k}'''}{(2\pi)^9}
\langle (\delta \chi_{\V{k}_1-\V{k}'} \delta \chi_{\V{k}'})
(\delta \chi_{\V{k}_2-\V{k}''} \delta \chi_{\V{k}''})
(\delta \chi_{\V{k}_3-\V{k}'''} \delta \chi_{\V{k}'''})\rangle\; ,
\nonumber\\
&=&-\gamma^3N'^3
(2\pi)^3 \delta^3 (\sum_i \vec{k}_i)
{1\over 8}{\gamma_{,\chi}^3 \over \gamma^3} \int{d^3 \V{k}' \over (2\pi)^3 } \left({(2\pi^2 \P)^3
\over k'^3
|\vec{k_1} +\vec{k'}|^3 |\vec{k_2} -\vec{k'}|^3 } + 7 \; \rm{perms}\right)\; ,\nonumber \\
&=&-\gamma^3N'^3
(2\pi)^3 \delta^3 (\sum_i \vec{k}_i)
{1\over 8}{\gamma_{,\chi}^3 \over \gamma^3} (2\pi^2 \P)^3 B(\vec{k_1},\vec{k_2},\vec{k_3})\; .
\eea
Now the loop integral involves two different momenta
\be\label{shape}
B(\vec{k_1},\vec{k_2},\vec{k_3}) = \int \frac{d^3\V{k}'}{(2\pi)^3}\left( {1 \over k'^3
|\vec{k_1} +\vec{k'}|^3 |\vec{k_2} -\vec{k'}|^3 } +  7 \; \rm{perms} \right)\; .
\ee
Diagrammatically this is equivalent to a triangular loop of scalars (see Fig.~(\ref{tria})).
We note that near the poles at $\vec{k}' = 0,\vec{k}_2,-\vec{k}_1$, we get logarithmic
divergences which are cut off by the IR scale $L$.
This logarithmic dependence breaks scale invariance.
So our shape $B$ is a function of three variables which we choose to simply be the
norm of all three vectors $k_1, k_2,k_3$. An estimate of the shape can be obtained by simply
evaluating the integral around each poles, cutting off the momentum integration in the infrared
at scale $1/L$. So for example, the integrand
\be
\int \frac{d^3 \V{k}'}{(2\pi)^3} {1 \over k'^3
|\vec{k_1} +\vec{k'}|^3 |\vec{k_2} -\vec{k'}|^3}
\ee
has a pole around $\vec{k'} = 0$, and the integrand falls off rapidly when $k'$ becomes
of the same order as $k_1$ or $k_2$. Hence we can approximate the integral around that pole as
\be
\int \frac{d^3 \V{k}'}{(2\pi)^3} {1 \over k'^3
|\vec{k_1} +\vec{k'}|^3 |\vec{k_2} -\vec{k'}|^3} = \frac{\ln(\rm{Min}(k_1,k_2) L )}{2\pi^2 k_1^3k_2^3}
+ \cdots \; .
\ee
The same thing can be done for the other poles and for the various permutations.
There are also points in parameter space where the integrand has a pole of order 4.
These poles occur in the squeezed limit where $\vec{k_1} = -\vec{k_2}$ and
hence $\vec{k}_3 \rightarrow \vec{0}$. This shows that the bispectrum diverges in the squeezed
limit, as is usual for the local shape. In principle, we can only measure $k$ to a resolution
$\sim 1/L$ and the bispectrum, while large, is finite and of order $L^3/(3 k_i^3)$ in this limit. Hence the
stronger poles that we have neglected are only important in the squeezed limit and they give contributions
of the same order as the log terms in that limit.
The full approximative shape is
\be\label{approximateshape}
B(k_1,k_2,k_3) \approx \frac{8}{2\pi^2}\left(\frac{\ln (\rm{Min}(k_1,k_2) L)+1/3}{ k_1^3k_2^3}
+ \rm{2\;\;perm.}\right)\; .
\ee
The $1/3$ term is only relevant for scales smaller than $k \sim e^{1/3} \frac{1}{L}$. For larger $k$
the shape is very well approximated by
\be
B(k_1,k_2,k_3) \approx \frac{8}{2\pi^2}\ln (\rm{Min}(k_i) L)\frac{\sum_i k_i^3}{\prod_ik_i^3}\; .
\ee
We show numerically in Appendix \ref{sectionshape} that this is a good approximation.
In Figure (\ref{shape}), we plotted the shape given by Eq.~(\ref{approximateshape}) in term of the 
usual variable
$x_2 = k_2/k_1$ and $x_3= k_3/k_1$. When the bispectrum is scale invariant, $k_1$ is fixed to 1 
(arbitrarily) but here we plotted
the shape for different value of $k_1$. As the figure clearly shows, the graph is very close to 
local and the magnitude grows as $k_1$ increases.
\begin{figure}[ht]
\centering
\includegraphics[width=6in]{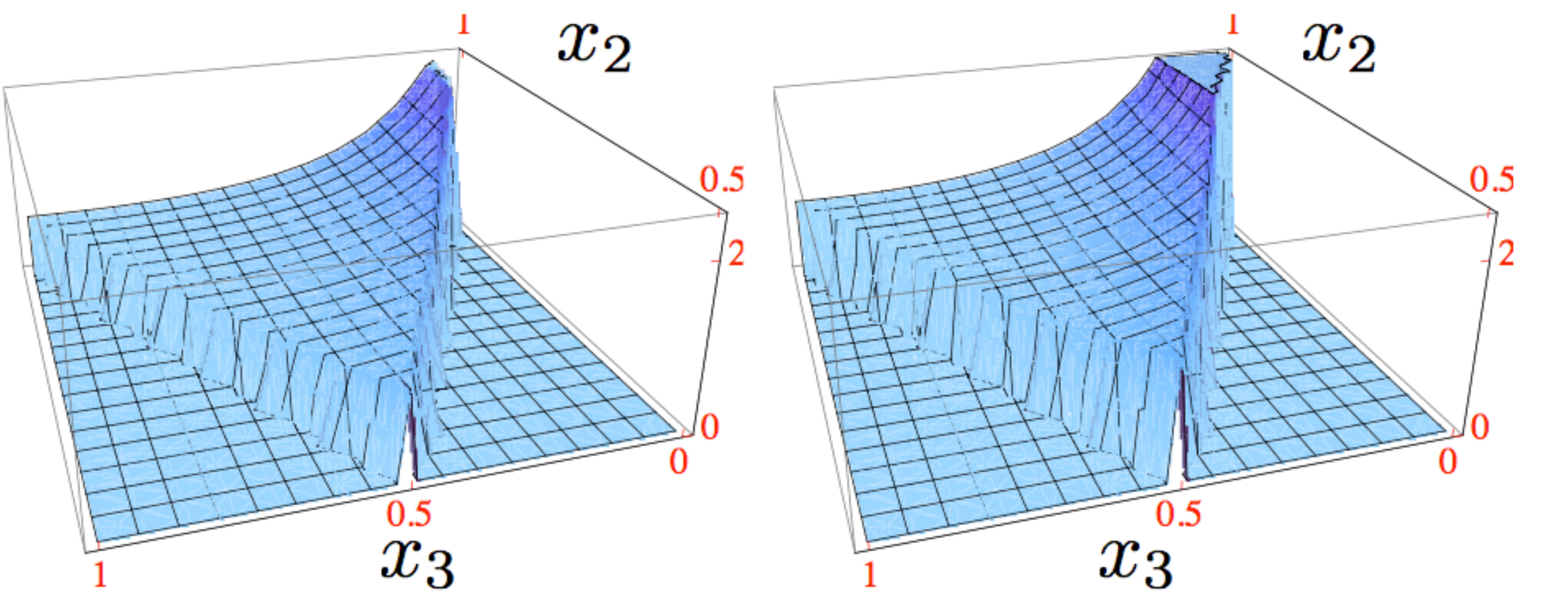}
\caption{Plot of the approximate shape $B(k_1,k_1x_2,k_1x_3) x_2^2 x_3^2 k_1^6$ (with $B(k_1,k_2,k_3)$ 
given by Eq.~(5.6)) in terms of $x_2 = \frac{k_2}{k_1}$ and $x_3= \frac{k_3}{k_1}$ for $k_1 = 0.5$ (left) 
and $k_1 = 1.5$ (right). The shape was restricted to be in the quadrant defined 
by $k_1 (1-x_2) < k_1 x_3 < k_1 x_2$ due to momentum conservation and to avoid overcounting 
identical triangle configurations (see \cite{Babich:2004gb}). The shape is clearly very close to 
local with the strongest signal in the squeezed limit when $k_3 = k_1 x_3 \rightarrow 0$. The 
overall magnitude of NG increases with the wavenumber $k_1$ or as we consider smaller wavelengths.}\label{shape}
\end{figure}
At the equilateral
point $k_1= k_2= k_3 \equiv k$, the loop contribution to the bispectrum simplifies to
\bea\label{equilateral}
\Expect{\zeta^3}&=&-\gamma^3 N'^3 (2\pi)^3
\delta^3 (\sum_i \vec{k}_i)
{\gamma_{,\chi}^3 \over \gamma^3}
\ln(kL) (2\pi^2)^2 \P^3 \frac{3}{k^6}\; .
\eea
If we compare the standard parameterization for local
non-Gaussianities (Eqns.~(\ref{localansatz}) and (\ref{fnldef2}))
at the equilateral point to Eq.~(\ref{bitree}) and Eq.~(\ref{equilateral})
and using the approximation $\P^\zeta \approx N'^2 \P$, we have
\be
f_{NL} \approx  - {5\over 6}  {\gamma^2 \gamma_{,\chi} \over N'}
\left(1 +\frac{\gamma_{,\chi}^2}{\gamma^2} \ln(kL) \P\right)\; ,
\ee
where the first term is the tree-level contribution, and the
second term is the one-loop contribution.

In the case of the two-point function, experimental bounds on the spectral index
required the loop-contribution to be subleading.  But there is
no such requirement for the bispectrum.  The loop contribution will
dominate if
\be
\frac{\gamma_{,\chi}^2}{\gamma^2} \P \ln(kL) >1\; .
\ee
In this limit we have
\bea
|f_{NL}| \approx   {5\over 6}  {(\gamma_{,\chi}^2 \P)^{3\over 2}
\over N' \P^{1\over 2}} \ln(kL) \lsim 100 \ln (kL)\; ,
\eea
where we have utilized the bound $\gamma_{,\chi}^2 \P < 10^{-2}$ and
the normalization $\P_\zeta^{1/2} \sim N'\P^{1/2} \sim 10^{-5}$
from COBE data.  We thus find, in this scenario, that one can
easily generate local non-Gaussianity which is not ruled out by
WMAP5 and can potentially be probed at Planck. Note that the magnitude of the non-Gaussianity
increases logarithmically with momentum, suggesting that non-Gaussianity
can have an important impact on the formation of structure at smaller
scales. If we define the running of $f_{NL}$ at the equilateral point
\be
n_{NG} =\left. \frac{d\ln f_{NL}}{d\ln k}\right|_{k_i = k}
\ee
one gets in the loop dominated limit
\be
n_{NG} \simeq {1\over \ln(kL)}\; .
\ee
In the limit where non-linearities dominate, the running of
$f_{NL}$ is thus independent of $N'$, $\gamma$
and $\gamma_{,\chi}$.

\subsection{Trispectrum}

As in the case of the 3-point function, the only non-vanishing contributions
will arise from the non-linear dependence of $\delta N$ on $\delta \chi$, so we
can ignore $\delta \phi$ fluctuations.
To simplify notation, we define
\be
\zeta_{\V{k}} = A \delta\chi_{\V{k}} +B
\int \frac{d^3\V{k}'}{(2\pi)^3} \delta\chi_{\V{k} - \V{k}'}\delta \chi_{\V{k}'}
-B\Expect{\delta\chi_{\V{k}}^2}\; ,
\ee
where $A = - N^\prime \gamma$ and $B= - \frac12N^\prime\gamma_{,\chi}$. The
last term ensures that we only keep the connected part of every diagrams.  The tree level
contribution (the term of lowest order in $B$) is
\bea
\Expect{\zeta_{\V{k}_1}\zeta_{\V{k}_2}\zeta_{\V{k}_3}\zeta_{\V{k}_4}} &=& A^2B^2
\int {d^3\V{k}'\over (2\pi)^3 } {d^3\V{k}'' \over (2\pi)^3}
\Expect{\dc_{\V{k}_1}\dc_{\V{k}_2}\dc_{\V{k}_3-\V{k}'}\dc_{\V{k}'}\dc_{\V{k}_4-\V{k}''}\dc_{\V{k}''}}
+ 5\;\rm{perm}\; ,
\nonumber\\
&=& (2\pi)^3 4 A^2B^2
\delta^3\left(\sum \vec{k_i}\right) \left[P(k_1+k_3)P(k_1)P(k_2)
+ 11\;\rm{perm}
\right]\; ,
\nonumber\\
&=& 4A^2B^2 (2\pi)^3 \delta^3\left(\sum \vec{k_i}\right) \frac{T(k_i)}{N'^6}\; ,
\eea
where we have used that $\P^\zeta \sim N'^2\P$ and the shape is given by
\be
T(k_i) = \left(\frac{(2\pi^2\P^\zeta)^3}{(k_1k_{13} k_2)^3} + 11\;\rm{perm}\right)\;
\ee
with the notation $k_{ij} = |\vec{k_i} + \vec{k_j}|$. The magnitude of the trispectrum is usually given
by two numbers ($\tau_{NL}$ and $g_{NL}$) corresponding to two
distinct shapes:
\bea
\Expect{\zeta^4} = (2\pi)^3\delta^3\left(\sum \vec{k_i}\right)  \left[\tau_{NL} T(k_i)
+{54\over 25} g_{NL} (P^\zeta(k_2) P^\zeta(k_3) P^\zeta(k_4) + 3\;\rm{perm})\right]\; .
\eea
The lowest order contribution thus corresponds to $g_{NL}=0$ and $\tau_{NL}=4A^2 B^2 / N'^6$.
The 1-loop contribution comes from the following term
\bea
\Expect{\zeta^4}_{1-loop} & = & B^4 \int {d^3\V{k}' \cdots d^3\V{k}^{iv} \over (2\pi)^{12}}
\Expect{\dc_{\V{k}_1-\V{k}'}\dc_{\V{k}'}\cdots \dc_{\V{k}_4-\V{k}^{iv}}\dc_{\V{k}^{iv}}}\; ,\\
& = & (2\pi)^3  (16) B^4 \delta^3\left(\sum \vec{k_i}\right)\left[
\int {d^3k' \over (2\pi)^3} \frac{(2\pi^2\P)^4}{k'^3 |\V{k}_1-\V{k}'|^3|\V{k}_1+
\V{k}_2-\V{k}'|^3|\V{k}_3+\V{k}'|^3}
+ 5\;\rm{perm}\right]\; .
\nonumber
\eea
The integral over momentum is difficult in general, so we will only
estimate its value at the equilateral point $|k_i| = k$
\be
\Expect{\zeta^4}_{1-loop} = 16 (2\pi)^3 B^4
\delta^3\left(\sum \vec{k_i}\right)
(2\pi^2\P) {\ln(kL)\over 2\pi^2 } \frac{T(k_i)}{N'^6}
\ee
and thus
\bea
\tau_{NL} &=& \frac{4B^2}{N'^6} \left( A^2 + 4 B^2\P \ln (kL)\right)\; ,
\nonumber\\
&=& \frac{\gamma^2 \gamma_{,\chi}^2}{N'^2} \left(1+
\frac{\gamma_{,\chi}^2}{\gamma^2} \P \ln(kL)\right)\; ,
\nonumber\\
g_{NL} &=& 0\; .
\eea
We see that the trispectrum is dominated by the non-linear contributions in largely the same
regime as the bispectrum.  Given the bound from $n_s-1$, the maximum value for $\tau_{NL}$ in this
loop dominated regime is
\bea
\tau_{NL} & \sim &\frac{\gamma_{,\chi}^4 \P ^2}{N'^2 \P} \ln(kL)
  <  10^6\ln(kL)\; .
\eea
Interestingly, the bound from WMAP5 on this parameter is $|\tau_{NL}|<10^8$ while Planck is expected
to improve this bound up to $|\tau_{NL}| < 560$.

\section{Conclusions}

We have studied a simple class of models in which non-Gaussianity is
dominantly produced by higher-order non-linearities in the
transfer of fluctuations
from the fundamental scalars to the curvature.  These higher-order non-linear
order contributions are often referred to in the literature as ``c-loops", and
can dominate the lowest order ``tree-level" contribution in the limit where
${\gamma_{,\chi}^2 \over \gamma^2} \P \ln(kL) >1$, where $\gamma$ and $\gamma_{,\chi}$
parameterize the non-linear transfer of fluctuations.
In particular, $f_{NL} \sim 100$ can be achieved in these models.

We have also found in these models that the magnitude of non-Gaussianity is
scale dependent, with
$n_{NG} \sim 0.2$ at CMB scales and $n_{NG} \sim 0.1$
at LSS scale.  Interestingly, the non-Gaussianity of the bispectrum is stronger at smaller
scales, where it can potentially be observed by large scale structure experiments.
The shape of our NG signal is very nearly local.
Moreover, this class of models yields a non-trivial trispectrum (parameterized by $\tau_{NL}$)
that also runs.

A number of open issues remain. In our model, we have assumed
that the slow-roll parameter $\epsilon$ is constant throughout inflation.
This was necessary in order to have an observable effect from the end of inflation,
but it requires tuning and it leads to a very flat power spectrum.
It would be interesting to either relax this assumption in our scenarios or to look
at a completely different set-up where the NG is not generated at the end of inflation.
We expect that we can relax this assumption since we could have a case where $f_{NL}$ is very small
on CMB scales but grows to be detectable on LSS scales.  We note though
that D-term inflation with a Coleman-Weinberg potential (as illustrated in Appendix A) has a natural regime
with the required flat potential, $\epsilon_e \sim \epsilon_f$.  From an effective field theory point of view
(and from string theory models such as \cite{Dutta:2007cr, Haack:2008yb}),
 the real tuning is in keeping all other allowed terms (such
as a mass term for $\phi$) subdominant to the Coleman-Weinberg potential. 

We have also assumed that the fundamental scalars ($\phi$ and $\chi$) are
Gaussian, and that all non-Gaussianity is induced by the non-linear
transfer of $\delta\chi$ fluctuations to the curvature.  Non-trivial NG can also arise
from non standard kinetic terms, or a steep potential for $\chi$ (which unlike the inflaton does not
have to satisfy slow-roll conditions). Loop corrections then have a richer structure although the basic idea remains the same. Of particular interest are models like DBI inflation where the spectral index is nearly one and entropy modes being converted to curvature at the end of inflation can also be observable \cite{Leblond:2006cc}.  This scenario
has been analyzed recently in \cite{RenauxPetel:2009sj} based on methods developed in \cite{Langlois:2008wt, Langlois:2008qf} (see also \cite{Gao:2009gd}\cite{Gao:2009fx}) and a mixture of equilateral and local NG has been found. It would be interesting to consider the regime where the loop dominate in this kind of models.

\acknowledgments
We are particularly thankful to Bhaskar Dutta for early collaboration on this project. We are grateful
to Niayesh Afshordi, Sarah Shandera, Martin Sloth, Xerxes Tata and Andrew Tolley for useful discussions.
L.L. would like to thank the organizers of the workshop on
Effective Field Theory of Inflation at the Perimeter Institute and of the Phenomenology workshop at
Cooks Branch Conservancy where part of this work was presented. L.L. would also like thank the KITP and
the Aspen Institute
for their hospitality.  LL is supported in part by  NSF Grant No.~PHY--0505757.
AR is supported in part by NSF Grant No.~PHY--0653656. This research was supported in part by Perimeter 
Institute for Theoretical Physics. Research at Perimeter Institute is supported by the Government of 
Canada though Industry Canada and by the province of Ontario through the Ministry of Research \& Innovation.
\appendix

\section{A Specific Model}\label{details}

The discussion in the text is very general and the ultimate goal of having dominant loop contribution
in the bispectrum inducing a large running may be achievable in a variety of ways. Here we we illustrate
the necessary ingredients with a specific model (based on \cite{Dutta:2008if}).
Take an inflationary potential
\bea
V_{\rm{inf}} & = & {g^2 \xi^2 \over 2}
\left[1+{g^2 \over 16\pi^2}V_{CW}(x)
\right]\; ,\\
V_{CW}(x) & = & (x^2+1)^2\ln(x^2+1)-2x^4\ln x^2+ (x^2-1)^2\ln(x^2-1) -4\ln2\; ,\nonumber
\eea
where $x^2 = \frac{\lambda^2 \phi^2}{g^2\xi}$, $\xi$ has mass dimension 2 and $\lambda$ and $g$
are dimensionless couplings. The reader will recognize this as the Coleman-Weinberg potential.
There is a regime in parameter space where the inflaton does not move very much
with
\be
\phi_*^2 \sim \phi_e^2 = \frac{g^2\xi}{\lambda^2}
\ee
and the slow-roll parameter is also nearly
constant\footnote{By integrating the EoM of motion of $\phi$, in the limit $x\rightarrow1$, one can
check that $\phi_* \sim \phi_e$ is a good approximation as long as $\frac{\xi}
{M_p^2 \lambda^2} \gg \frac{ 2\sqrt{2}\ln2 N_e}{\pi^2}$ where $N_e$ is the number of efolds
between horizon crossing and the surface of reheating.}
\be
\epsilon^\phi = \left(\frac{g^2 \ln 2}{\pi^2 \phi}\right)^2\frac{M_p^2}{2} \sim \frac{\lambda^2g^2
(\ln 2)^2 M_p^2}{2\pi^4\xi}\; .
\ee
The $\phi$ and $\chi$ power spectrum are simply given by $\P = \frac{H^2}{(2\pi)^2}$ and
they will remain approximately
constant until the end of inflation if $\eta_\phi$ and $\eta_\chi$ are much smaller than 1.
Now consider a simple potential for $\chi$
\be
V_{\rm{hid}} = \nu^2\chi^4/4\; .
\ee
This potential drives $\chi$ to $0$ but the field will fluctuate and acquire some stochastic value
$\overline{\chi}$ which in general will be non-zero (although small). The tachyon potential is of the
form Eq.~(\ref{Vmess}) with the surface of reheating defined by
\be
0=f(\phi_e,\chi_e) = -g^2\xi + \lambda^2\phi_e^2 + \beta\chi_e^2\; .
\ee
We choose a model such that
$\chi_e = \overline{\chi}_e + \delta\chi \ll \phi_e$.
Since the function $f$ is quadratic in both fields, the transfer function is simply
\bea
\gamma &=&\left.\frac{\pa \phi_e}{\pa \chi}\right|_e
= -{\beta \over \lambda^2}
{\overline{\chi}_e \over \phi_e}\; ,
\eea
and $\gamma_{,\chi} \sim \frac{\gamma}{\chi}$ while $\gamma_{,\chi\chi} \sim 0$. Note that $\gamma$ 
and $\gamma_{,\chi}$ can both be either sign depending on $\beta$.
The curvature power spectrum is (from Eq.~(\ref{powerspectrum}))
\be
\P_\zeta = \frac{H^2}{2(2\pi)^2 \epsilon^\phi M_p^2}\left(1+\gamma^2 +
\frac{\gamma_{,\chi}^2 H^2 \ln kL}{(2\pi)^2}\right)\; .
\ee
We are interested in the regime where
\be
1> \frac{\gamma_{,\chi}^2 H^2 \ln kL}{(2\pi)^2}> \gamma^2
\ee
and where the power spectrum $\P^\zeta$,  $f_{NL}$ and $\tau_{NL}$ are well approximated by
\bea
\P_\zeta & \sim &\frac{\pi^2}{6}\frac{\xi^3}{(2\ln 2)^2\lambda^2M_p^6}\; ,\nonumber\\
f_{NL} & = & \frac{5 \ln 2 \beta^3  H^2 M_p^2\ln kL}{6\pi^2\lambda^2g^2(2\pi)^2\xi^2}\; ,\nonumber\\
\tau_{NL} & = & \frac{\beta^4 M_p^4 H^2 \ln kL}{8\pi^4 (2\pi)^2\lambda^2g^2\xi^3}\; .
\eea
Note that none of these observables depend on the precise value of $\bar\chi$ at the end of inflation 
although in order for the loop contribution it must be that the average value of the zero mode of $\chi$ 
at the end of inflation is smaller than $H$.
We show a point in parameter space (see Table (\ref{parameters}))
where all the conditions mentioned in this section are respected.
\begin{table}[h]
\center
\begin{tabular}{|c|c|c|c|}\hline
$g$ & $\lambda$ & $\beta$ & $\xi/M_p^2$ \\ \hline
$ 10^{-3}$ & $2\times 10^{-4}$ &  $0.1$ & $2.98 \times 10^{-6}$ \\ \hline
\end{tabular}
\caption{A point in parameter space. $\lambda$ was first chosen and $\xi$ was solved for by
matching to COBE data. The parameter $g$ is then constrained such that $\phi < M_p$ by at least
two order of magnitude. $\beta$ is a free parameters that
determine the magnitude of NG. The stochastic value of $\chi$ at the end of inflation in this model 
is less than H although none of the observables depend on its precise value. One can check that for 
this choice of parameters:
$\lambda^2\phi_e \gg \lambda'\chi^2$, $\frac{\gamma_{,\chi}^2 H^2 \ln kL}{(2\pi)^2} = 0.03$ and
$\gamma^2 \sim 0.001$. Also for this choice of parameter the potential is very flat with
$\epsilon_*^\phi \sim \epsilon_e^\phi \sim 10^{-11}$ while $\epsilon^\chi_e \sim 10^{-29}$ 
(for $\nu^2 \sim 10^{-2}$ giving $\eta_\chi \sim 10^{-2}$). }\label{parameters}
\end{table}

\begin{table}[h]
\center
\begin{tabular}{|c|c|c|c|c|c|c|}\hline
& $\ln kL$ & $\P_\zeta$ & $f_{NL}$ & $\tau_{NL}$ & $n_s -1 $ & $n_{NG}$ \\ \hline
CMB scales & $\sim 5$ & $2\times 10^{-9}$ & $1.6$ & 1152 & $0.004$ & $0.2$\\ \hline
LSS scales & $\sim 10$ & $2\times 10^{-9}$ &  $ 62$ & $4.5\times 10^4$& $0.004$ & $ 0.1$ \\ \hline
\end{tabular}
\caption{Predicted value for various parameters. This point was deliberately chosen to illustrate the
possibility of having a non-observable level of NG at CMB scale but with a very detectable signal
for LSS.}\label{predictions}
\end{table}

As can be seen from Table (\ref{predictions}), this simple model can lead to interesting observational
signatures whereas the CMB is very Gaussian, but significant NG appears for large scale structure. On
the other hand, not everything is perfect since the spectral index is nearly one in tension with the
most current WMAP5 data. This is direct consequence of working with a model where
$\epsilon_*^\phi \sim \epsilon_e^\phi$. If this assumption is relaxed, a running will be induced
but the non-Gaussian signal coming from the end of inflation will also be reduced. This can potentially
be compensated by varying other parameters but this required more detailed analysis keeping track of the
time of evaluation for each quantity. We leave this for further work.  Note also that the tension
between $n_s \sim 1$ and WMAP5 data can also be reduced if cosmic strings (which are generically produce
in these hybrid models) contribute to the density perturbation spectrum~\cite{Bevis:2007gh}.  The cosmic
strings will add their own source of NG
which will further constrain the model \cite{Hindmarsh:2009qk}.

\section{Numerical Evaluation of the Integral}\label{sectionshape}
The integral
\be
B(\vec{k_1},\vec{k_2},\vec{k_3}) = \int \frac{d^3 k'}{(2\pi)^3}\left( {1 \over k'^3
|\vec{k_1} +\vec{k'}|^3 |\vec{k_2} -\vec{k'}|^3 } +  7 \; \rm{perms} \right)
\ee
can be evaluated numerically in Mathematica.  There are three poles in the integrand,
and the IR cutoff discussed in Sec. (\ref{IR}) is most easily implemented by setting the
integrand to zero whenever $k'$ is within $1/L$ of a pole.

On general grounds, one expects that the integral is can be written as
\bea
B(\vec{k_1},\vec{k_2},\vec{k_3}) = \sum_i \ln(k_i L) F_0^i(|\vec{k_1}|,|\vec{k_2}|)
+ F_1(|\vec{k_1}|,|\vec{k_2}|) + \dots \; ,
\eea
where the neglected terms contain powers of $1/k_i L$.  For momenta relevant to
CMB, these terms are very small.  Note that both $F_0$ and $F_1$ are homogeneous
Lorentz invariant functions of the
external momenta with degree -6 (i.e., are scale-invariant), and
thus are completely determined by the norms of any two of the momenta.

As argued in Sec. (\ref{bispectrum}), the leading
term is dominated by the poles of the integrand, and we expect it to be of the
form
\bea
\sum_i \ln(k_i L) F_0^i (\vec{k_1},\vec{k_2},\vec{k_3}) =
\frac{8}{2\pi^2}
\frac{\sum_i \ln (\rm{Min}(k_{\alpha \neq i}) L)
k_i^3}{\left( \prod_j k_j^3 \right)}\; .
\eea
Because $F_{0,1}$ are scale-invariant,
we can numerically integrate the shape at various scales and fit the
result to our ansatz in order to determine the magnitude of $F_{0,1}$.

For example, we numerically integrated $B$ at the equilateral limit
$k_i = k$ for various values of $k$ near the range $kL \approx 150$,
and we fitted the results to the ansatz $B(k,k,k) \times (kL)^6 = c_0
\ln (kL) + c_1$.  This fit yielded the expected $c_0 \sim {12\over \pi^2}$,
where the second term  $c_1$ was $\sim 5\%$ of the first term.
Similarly, one can integrate $B$ for several external momenta in the squeezed limit where
$k_1 L= k_{small}L \sim 150$ and $k_2L = k_3 L = k_{big}L \sim 10000$.
Since the local shape should dominate in this limit, one would fit
this to $B(k_{small},k_{big},k_{big})\times (k_{small}L)^3 (k_{big}L)^3
\sim d_0 \ln(k_{small}L)  + d_1$.  Again as expected, one finds
$d_0 \sim {8\over \pi^2}$, where the non-logarithmic term $d_1$
was $\sim 8\%$ of the logarithmic term.  We conclude that the
logarithm captures the leading behavior.

\providecommand{\href}[2]{#2}\begingroup\raggedright\endgroup

\end{document}